\documentclass[twocolumn]{aastex61}
\usepackage{amsmath}

\submitjournal{ApJ (Accepted)}

\shorttitle{Pair flow and current composition in the pulsar magnetosphere}
\shortauthors{Brambilla et al.}

\begin{document}

\title{Electron positron pair flow and current composition in the pulsar magnetosphere}

\correspondingauthor{Gabriele Brambilla}
\email{gb.gabrielebrambilla@gmail.com}

\author[0000-0002-3692-1974]{Gabriele Brambilla}
\affiliation{Dipartimento di Fisica, Universit\`a degli Studi di Milano, Via Celoria 16, 20133 Milano, Italy}
\affiliation{Astrophysics Science Division, NASA Goddard Space Flight Center, Greenbelt, MD 20771, USA}
\affiliation{Istituto Nazionale di Fisica Nucleare, sezione di Milano, Via Celoria 16, 20133 Milano, Italy}

\author{Constantinos Kalapotharakos}
\affiliation{Astrophysics Science Division, NASA Goddard Space Flight Center, Greenbelt, MD 20771, USA}
\affiliation{University of Maryland, College Park (UMCP/CRESST), College Park, MD 20742, USA}

\author{Andrey N. Timokhin}
\affiliation{Astrophysics Science Division, NASA Goddard Space Flight Center, Greenbelt, MD 20771, USA}
\affiliation{University of Maryland, College Park (UMCP/CRESST), College Park, MD 20742, USA}

\author{Alice K. Harding}
\affiliation{Astrophysics Science Division, NASA Goddard Space Flight Center, Greenbelt, MD 20771, USA}

\author{Demosthenes Kazanas}
\affiliation{Astrophysics Science Division, NASA Goddard Space Flight Center, Greenbelt, MD 20771, USA}

\begin{abstract}
We performed ab-initio Particle-In-Cell (PIC) simulations of a pulsar magnetosphere with electron-positron plasma produced
only in the regions close to the neutron star surface.  We study how the magnetosphere transitions from the vacuum to a nearly force-free configuration. We compare the resulting force-free like configuration with ones obtained in a PIC simulation where 
particles are injected everywhere as well as with macroscopic force-free simulations. We found that although both PIC solutions have similar structure of electromagnetic fields and current density distributions, they have different particle density distribution. 
In fact in the injection from the surface solution, electrons and positrons counterstream only along parts of the return current regions and most of the particles leave the magnetosphere without returning to the star.  We also found that pair production in the outer magnetosphere is not critical for filling the whole magnetosphere with plasma.
We study how the current density distribution supporting the global electromagnetic configuration is formed by analyzing particle trajectories. We found that electrons precipitate to the return current layer inside the light cylinder and positrons precipitate to the current sheet outside the light cylinder by crossing magnetic field lines contributing to the charge density distribution required by the global electrodynamics.  Moreover, there is a population of electrons trapped in the region close to the Y-point.  On the other hand the most energetic positrons are accelerated close to the Y-point. These processes can have observational signatures that, with further modeling efforts, would help to distinguish this particular magnetosphere configuration from others.

\end{abstract}


\section{Introduction}{\label{sec:intro}}
%%%%%%%%%%%%%%%%%%%%%%%%%%%%%%%%%%%%%%%%%%%%%%%%%%%%%%%%%%%%%%%%%
In the last decade, large steps have been taken in the understanding of the pulsar magnetosphere problem on both theoretical and observational sides. On the observational side the launch of the Fermi $\gamma$-ray Space Telescope \citep{Fermi} led to the detection of 205 $\gamma$-ray pulsars (updated to 2016 February 22). On the theoretical side, the increased power of contemporary computers allowed verification and exploration of theoretical ideas with computationally expensive techniques. Global magnetospheric solutions have been obtained in ideal force-free electrodynamics (e.g. \citealt{CKF1999}; \citealt{Spitkovsky06}), dissipative electrodynamics (e.g. \citealt{Li12}; \citealt{Kala12}; \citealt{Kala14}; \citealt{Kala17a}), and then with particle-in-cell simulations (PIC, \citealt{Philippov14}; \citealt{Chen2014}; \citealt{Cerutti2015}; \citealt{Philippov15a}; \citealt{Belyaev2015a}; \citealt{Belyaev2015b}; \citealt{Philippov15b}; \citealt{Cerutti2016}; \citealt{Cer16b}; \citealt{Philippov17}; \citealt{Kalapotharakos2017b}). 
PIC simulations can simulate the pulsar magnetosphere from first principles, in contrast with dissipative electrodynamics simulations. The previous PIC works with pulsar magnetospheres addressed important problems such as the global magnetospheric currents and their composition, the dissipative processes and the electromagnetic emission, the role of pair production and of general relativity. In this paper, we focus on the magnetospheric structure that arises when particles are supplied only at the neutron star surface. We also inject a larger range of pair multiplicity than previous works which makes the simulation more realistic and allows us to explore the dependence of magnetosphere properties on particle injection rate. 
\newline
This paper is structured as follows: in \S\,\ref{sec:methods}, we present the
simulation setup; in
\S\,\ref{sec:magn_screen}, we present how the magnetosphere transitions from vacuum to the force-free limit injecting particles only from the surface, in \S\,\ref{sec:ff_solutions}, we compare the force-free solution obtained injecting particles from the surface to the one obtained injecting particles everywhere. We compare both macroscopic quantities and particle trajectories. In \S\,\ref{sec:discussion} we discuss our findings; \S\,\ref{sec:conclusions} is the conclusion with a summary of the results and outlooks.
%%%%%%%%%%%%%%%%%%%%%%%%%%%%%%%%%%%%%%%%%%%%%%%%%%%%%%%%%%%%%%%%%%
\section{Simulation setup and methods}{\label{sec:methods}}
%%%%%%%%%%%%%%%%%%%%%%%%%%%%%%%%%%%%%%%%%%%%%%%%%%%%%%%%%%%%%%%%%%
The explicit electromagnetic PIC technique \citep{birdsall} works in a cycle, with the electromagnetic fields pushing some particles representing the plasma (usually called macro particles) determining the current that later enter as sources in the Maxwell equations to modify the fields. 
The simulations presented in this paper are obtained with the electromagnetic and relativistic PIC code C-3PA presented in \cite{Kalapotharakos2017b}, where simulations with particles injected everywhere in the domain are discussed. The only major difference is that we modified the code in order to save separately the contribution of the two particle species to the currents:
\begin{equation}
\mathbf{J}=\mathbf{J_{pos}}+\mathbf{J_{ele}}
\end{equation}
where $\mathbf{J_{pos}}$ and $\mathbf{J_{ele}}$ represent the positron and electron currents, respectively. We will use these quantities mainly in section \ref{sec:macroscopic}.\newline
Here we want to analyze simulations of the pulsar magnetosphere with pair plasma supplied only at the surface of the star.
The neutron star has effective radius of $0.36 R_{\rm LC}$ ($R_{\rm LC}=c/\Omega$ is the light cylinder, where $c$ is the speed of light and $\Omega$ the angular frequency of the neutron star) with conductive boundary conditions implemented below $0.28 R_{\rm LC}$ and a kernel layer between $0.28 R_{\rm LC}$ and $0.36 R_{\rm LC}$ \citep{Kalapotharakos2017b}.
We inject particles according to the local magnetization in each cell, $\sigma_M = B^2/8\pi nm_ec^2$ ($B$ the magnetic field, $n$ the number density, $m_e$ the electron mass), rather than at a fixed rate. Regulating the injection based on local magnetization prevents over-injection in the closed field regions where particles are trapped and the their number density increases more quickly than in the open field regions. Moreover, in our simulations we check that $\sigma_M > 10$ everywhere except in the pulsar current sheet, where the magnetic energy is converted into particle kinetic energy. This is necessary in order to study a well magnetized plasma, such as in a real pulsar magnetosphere. Such an injection with respect to magnetization helps achieve this result.
Particles are injected until the local magnetization $\sigma_M$ in each cell is below the threshold:
\begin{equation}
\Sigma = \Sigma_0\left(\frac{r_0}{r}\right)^3
\end{equation}
Where $r_0$ is the stellar radius and $r$ is the radial coordinate in spherical coordinates.  $\Sigma_0$ is a quantity that is globally known in the simulation and that is assigned an initial value (which is, for example 2400). The particle supply is regulated to achieve a prescribed global injection rate, $\mathcal{F}$. Every 10 time steps, we count all the particles injected at that time step and we compare this rate to $\mathcal{F}$. If this rate is lower/higher than $\mathcal{F}$, we decrease/increase $\Sigma_0$. It takes less than a third of a stellar rotation to achieve a desired $\mathcal{F}$.
The unit of measure of $\mathcal{F}$ is:
\begin{equation}
\mathcal{F}_{\rm GJ} = \frac{2\rho_{\rm GJ}A_{pc}c}{q_e}
\end{equation}
where $\rho_{\rm GJ}$ is the Goldreich-Julian charge density in the pulsar polar cap \citep{GJ69},
\begin{equation}\label{eq:rhoGJapp}
\rho_{\rm GJ}\sim\frac{\Omega B_0\cos\alpha}{2\pi c}
\end{equation}
where $\alpha$ is the inclination angle between the rotation axis and the magnetic moment, $B_0$ is the magnetic field at the star pole, $q_e$ is the electron charge, and
\begin{equation}
A_{pc}\sim\pi r_0^3\Omega/c
%A_{pc}\sim\pi\left( r_0(r_0\Omega/c)^{1/2} \right)^2
\end{equation}
is the area of the polar cap. The factor $2$ accounts for the two poles. We introduce also the unit $\mathcal{F}^0_{\rm GJ}=\mathcal{F}_{\rm GJ}/\cos\alpha$ representing the $\rho_{\rm GJ}$ for an aligned rotator. Pairs are injected with zero velocity. \\
The simulation domains are cubes of side $9.6 R_{\rm LC}$ with the neutron star rotating at the center. A perfectly matched layer (PML) is implemented at the outer boundary of the domain \citep{PML1,PML2,KALAPOTHARAKOSandCONTOPOULOS2009}.  The main limitations of explicit PIC algorithms are the (temporal) resolution of the plasma frequency $\omega_p$ and the (spatial) resolution of the skin depth $\lambda_{sd}$. Not resolving $\omega_p$ generates a numerical instability, while not resolving $\lambda_{sd}$ causes numerical heating in the plasma. We use a grid size of $d=0.02 R_{\rm LC}$ and a time step small enough to resolve $\omega_p$ everywhere in the domain with at least three time steps ($dt=0.003 R_{\rm LC}/c$). We do not resolve $\lambda_{sd}$ approximately in a sphere of radius $0.9 R_{\rm LC}$ centered on the star. We observe that the numerical heating gives a small Lorentz factor to the particles. Even though this heating could push the particle energies up to $\gamma\approx 30$, we find that only 1.7\% of the particle population reach $\gamma>20$. The simulations in this paper are obtained with a $\gamma_{max}\sim500$
\begin{equation}
\gamma_{max}\sim\frac{\Omega^2 r_0^3 B_0 e}{m_e c^2}
\label{eq:openvolt}
\end{equation}
where $\Omega$ is the angular frequency of the neutron star rotation, $r_0$ the stellar radius, $B_0$ the magnetic field at the star radius. $\gamma_{max}$ is the Lorentz factor of an electron accelerated through all the voltage between the center of the polar cap of an aligned rotator and the last open field line (e.g. \citealt{RudSut75}). The value of $B_0$ and then $\gamma_{max}$ is necessary to resolve the plasma frequency in the simulation, because a realistic value of $B_0$ would increase the characteristic charge density at the star surface (Equation \ref{eq:rhoGJapp}), therefore, the time and space resolution needed would not be computationally accessible. The particles in our simulations are subjected to an enhanced radiation reaction to ensure a rapid cooling of the perpendicular momentum \citep{Kalapotharakos2017b}. Low $B_0$ and small $dt$ guarantee that the gyro frequency is resolved everywhere in the simulation. If not stated otherwise, when we show 2D slices of our simulations they are in the $\mathbf{\pmb \mu}$ - $\mathbf{\Omega}$ plane, where $\mathbf{\pmb\mu}$ is the magnetic moment and $\mathbf{\Omega}$  is the angular velocity that lies on the rotational axis. 
All the visualizations are obtained with VisIt \citep{VisIt}.
%%%%%%%%%%%%%%%%%%%%%%%%%%%%%%%%%%%%%%%%%%%%%%%%%%%%%%%%%%%%%%%%%%
\section{Formation of a force-free like magnetosphere}\label{sec:magn_screen}
%%%%%%%%%%%%%%%%%%%%%%%%%%%%%%%%%%%%%%%%%%%%%%%%%%%%%%%%%%%%%%%%%%
The possibility to fill the entire magnetosphere and make it nearly force-free everywhere injecting particles only from the surface has been shown in \cite{Cerutti2016}. However, \cite{Cerutti2016} focused mostly on the high-energy emission and not on the magnetosphere structure and its dependence on injection rate. As we increase the injection rate at the stellar surface we expect to find many different magnetosphere configurations ranging 
from charge-separated magnetospheres \citep{KP-Michel1985,Spitkovsky&Arons2002}
to a close to force-free solution \citep{CKF1999,Spitkovsky06}.  
%%%%%%%%%%%%%%%%%%%%%%%%%%%%%%%%%%%%%%%%%%%%%%%%%%%%%%%%%%%%%%%%%%
\subsection{Electromagnetic energy and Poynting flux}\label{sec:energy}
\begin{figure}
  \singlespace
  %\centering
  %\hspace{-.5in}
  \includegraphics[width=0.5\textwidth]{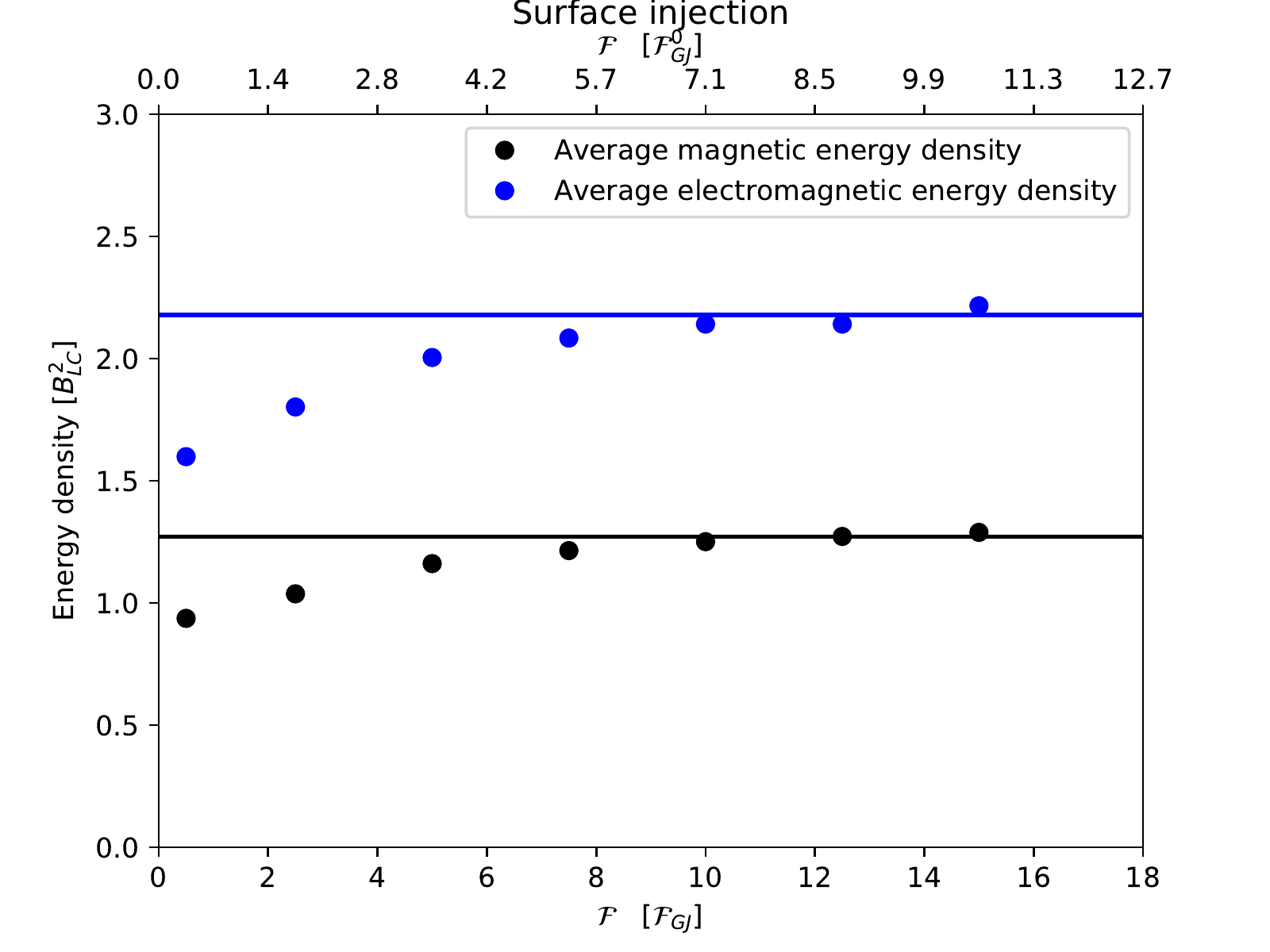}
  %\vspace{-0.2in}
  \caption{Average electromagnetic and magnetic energy density as a function of injection rate, for simulations with injection everywhere. The solid lines are the expected values from force-free electrodynamics. These kinds of simulations are the subject of the paper \cite{Kalapotharakos2017b}.}
  \label{fig:who-inj-fig}
\end{figure}

\begin{figure}
  \singlespace
  %\centering
  %\hspace{-.5in}
  \includegraphics[width=0.5\textwidth]{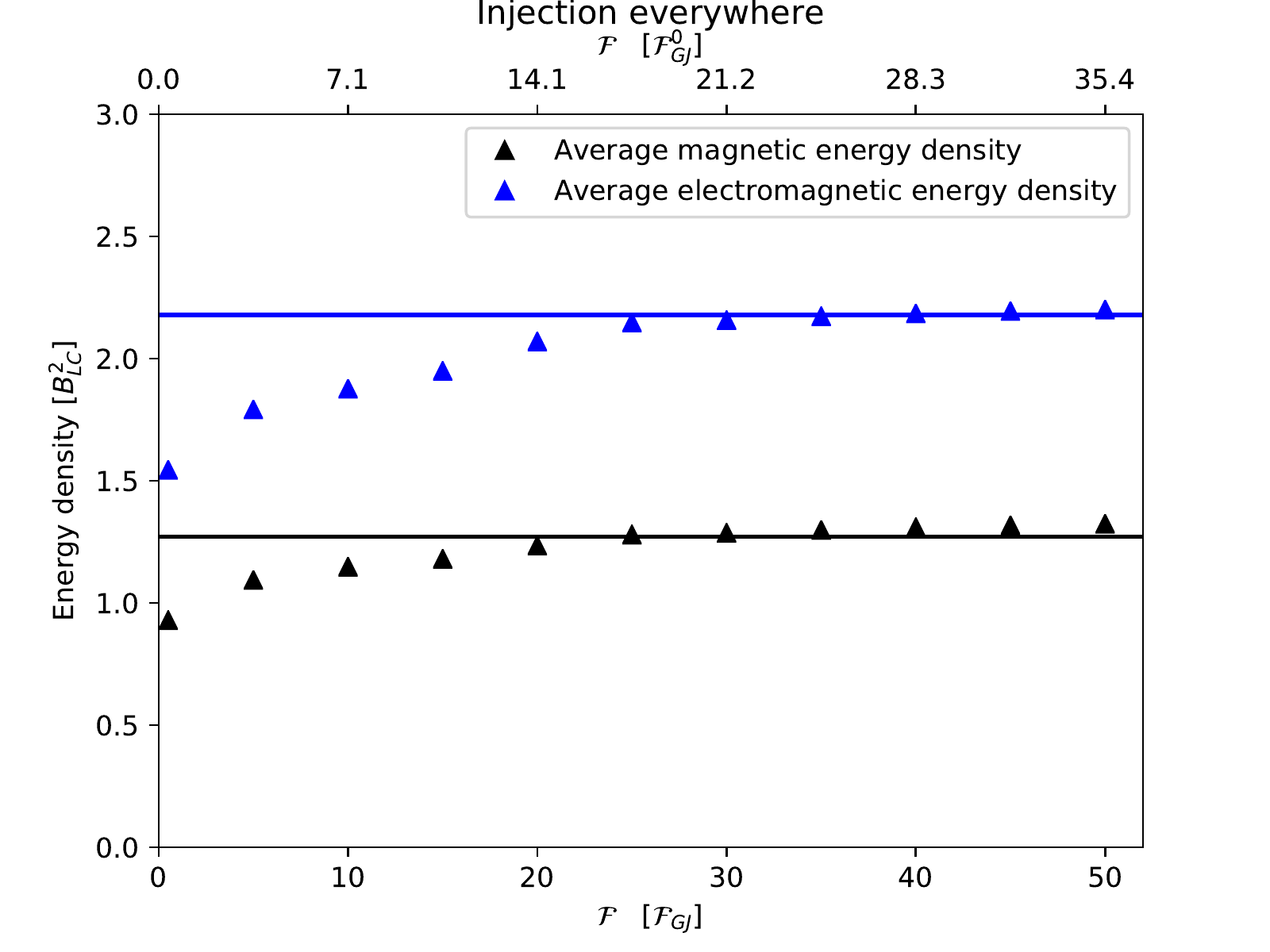}
  %\vspace{-0.2in}
  \caption{Average electromagnetic and magnetic energy density as a function of injection rate, for surface injection simulations. The solid lines are the expected values from force-free electrodynamics. }
  \label{fig:sur-inj-fig}
\end{figure}

\begin{figure*}
  \singlespace
  \centering
  \hspace{-.5in}
  \includegraphics[width=0.8\textwidth]{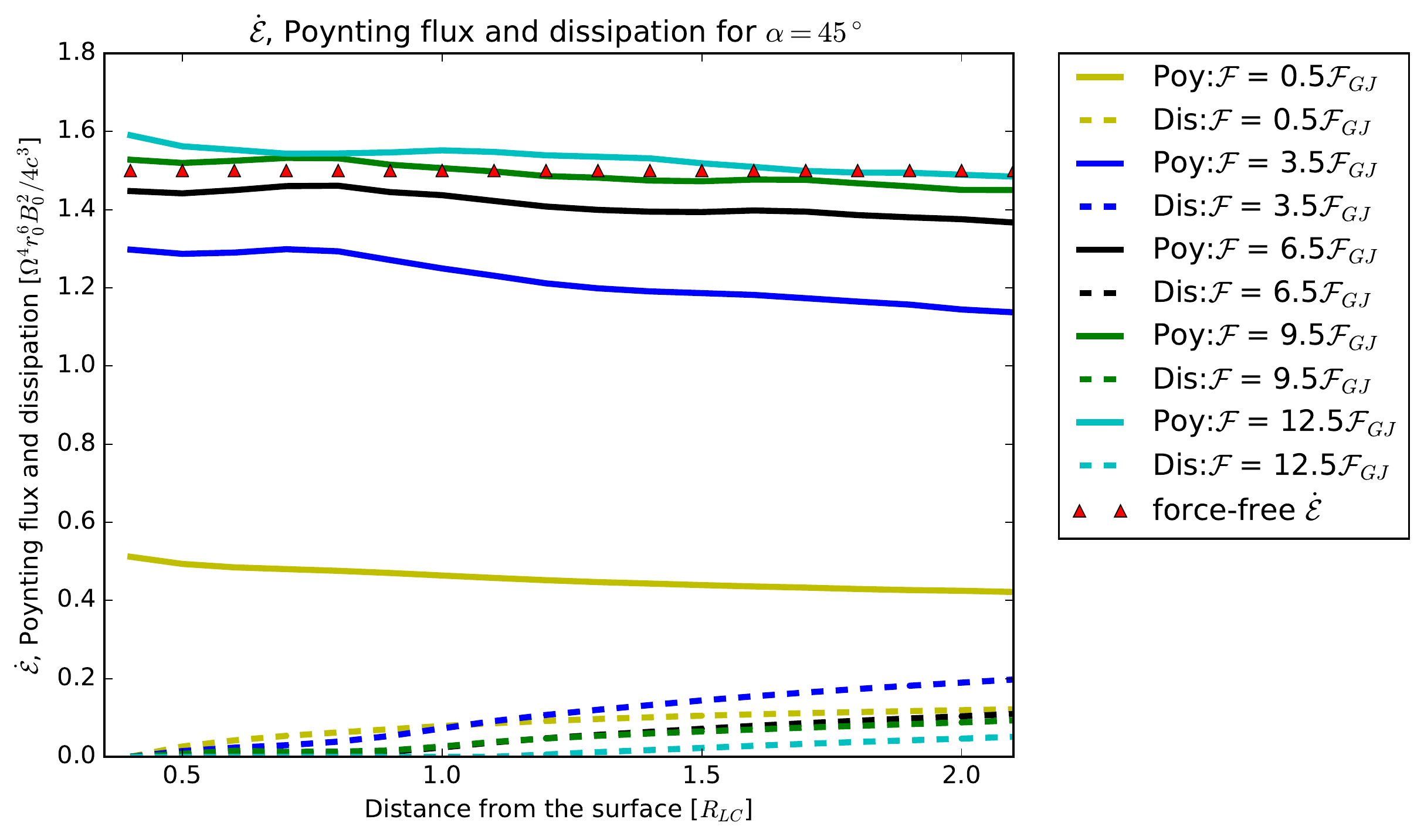}
  %\vspace{-0.2in}
  \caption{Variation of the Poynting flux and of the dissipation with the radial distance for different injection rates. This plot is for the $\alpha=45^{\circ}$ case. }
  \label{fig:poynting}
\end{figure*}

%%%%%%%%%%%%%%%%%%%%%%%%%%%%%%%%%%%%%%%%%%%%%%%%%%%%%%%%%%%%%%%%%%
An important question we try to answer is how close the electromagnetic field structure of a magnetosphere is to the one of the force-free limit. The force-free configuration is characterized by the value of energy stored in 
the electromagnetic fields (see for example \citealt{Bellan}). In each simulation we evaluated the average of the electromagnetic field energy density over the volume of a spherical shell starting outside the boundary layer of the star and extending up to $2.5~R_{\rm LC}$. We compared this result with the same quantity
evaluated in a macroscopic force-free electrodynamics simulation \citep{Kala12}.
We found that our global PIC models presenting both the Poynting flux close to the theoretical value and the electromagnetic field energy density close to the 
force-free electrodynamics simulations provide the current and charge density distribution and field structure close to the ideal force-free ones. 
We tested this method to the simulations with injection everywhere and then applied it to the simulations with injection from the surface. The simulations that inject particles over the entire computational domain showed that the average electromagnetic energy density value remains constant over a broad range of injection rates after reaching the force-free value. In Figures \ref{fig:who-inj-fig} and \ref{fig:sur-inj-fig} we show the average electromagnetic energy density versus the injection rate for pulsars with $\alpha=45^{\circ}$. We plot also the average magnetic energy density, however the two quantities show a very similar behavior.
In this paper we focus on magnetospheres obtained injecting particles close to the star surface and we tested the validity of the method described above to find close to force-free solutions over the whole range of inclination angles.
\begin{figure*}
  \singlespace
  \centering
  %\hspace{-.5in}
  \includegraphics[width=1.\textwidth]{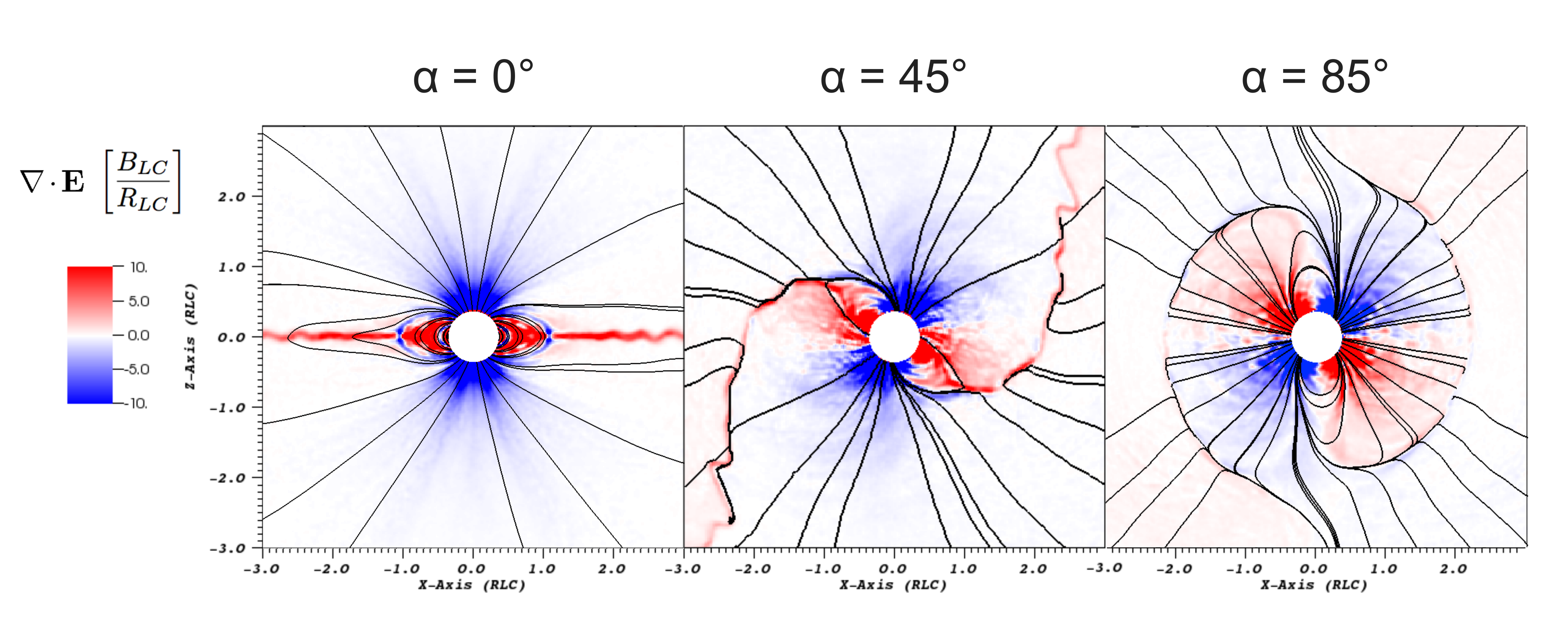}
  %\vspace{-0.2in}
  \vspace{-0.1in}
  \caption{In this figure we show the close to force-free simulations obtained injecting particles from the surface. We show cases that cover the whole range of inclination angles $\alpha$. The color is the divergence of the electric field, while the field lines are the magnetic field lines projected onto the poloidal plane. The $0^\circ$ case is obtained with $\mathcal{F}=5 \mathcal{F}_{\rm GJ}=5 \mathcal{F}^0_{\rm GJ}$, the $45^\circ$ case with $\mathcal{F}=12.5 \mathcal{F}_{\rm GJ}=8.84\mathcal{F}^0_{\rm GJ}$ and the $85^\circ$ with $\mathcal{F}=125 \mathcal{F}_{\rm GJ}=10.89\mathcal{F}^0_{\rm GJ}$.}
  \label{fig:diffANGLES}
\end{figure*}
In Figure \ref{fig:diffANGLES} we show the charge density of near force-free magnetospheres with inclination angle $0^\circ$, $45^\circ$, $85^\circ$ obtained with this method. We use $85^\circ$ instead of $90^\circ$, since for $\alpha=90^\circ$ the injection rate unit is $0$ (Equation \ref{eq:rhoGJapp}).  
The Poynting flux increases toward the force-free value as the injection rate increases. The spin down power $\dot{\mathcal{E}}$ of a rotating magnetized sphere in vacuum \citep{Deutsch1955} is
\begin{equation}\label{eq:vac-spin}
\dot{\mathcal{E}}=\frac{\Omega^4r_0^6B_0^2}{6c^3}\sin^2\alpha
\end{equation}
The empirical expression of $\dot{\mathcal{E}}$ for a force-free magnetosphere is \citep{Spitkovsky06}
\begin{equation}\label{eq:ff-spin}
\dot{\mathcal{E}}=\frac{\Omega^4r_0^6B_0^2}{4c^3}(1+\sin^2\alpha)
\end{equation}
In both these cases, $\dot{\mathcal{E}}$ is totally carried away by the Poynting flux. 
In Figure \ref{fig:poynting} we show the Poynting flux evaluated through spherical shells for pulsars with $\alpha=45^{\circ}$ with increasing injection rate. The Poynting flux decreases along the radial direction because it gets dissipated by $J \cdot E$ in the volume enclosed by the shell. However, we note that the dissipation, especially for the higher injection rates, takes place close to the $R_{\rm LC}$, and beyond that near the equatorial current sheet. 
In Figure \ref{fig:poynting} we see that the maximum dissipation happens for $\mathcal{F}=3.5\mathcal{F}_{\rm GJ}$ and does not exceed 15\% of the Poynting flux at the surface. The maximum percentage in dissipation happens for $\mathcal{F}=0.5\mathcal{F}_{\rm GJ}$ ($\sim$ 20\%) but this is because the absolute value of the Poynting flux is lower. 
The dissipation decreases monotonically going toward the force free solution from $\mathcal{F}=3.5\mathcal{F}_{\rm GJ}$. The dissipation for the nearest to force-free solution is $\sim$ 6\%. These results are consistent with the $\gamma$-ray efficiency in \cite{Kalapotharakos2017b} for magnetospheres with particles supplied everywhere.

%%%%%%%%%%%%%%%%%%%%%%%%%%%%%%%%%%%%%%%%%%%%%%%%%%%%%%%%%%%%%%%%%%
\subsection{Screening of the accelerating electric fields}\label{sec:screening}
\begin{figure*}
  \singlespace
  \centering
  %\hspace{-.5in}
  \includegraphics[width=1.\textwidth]{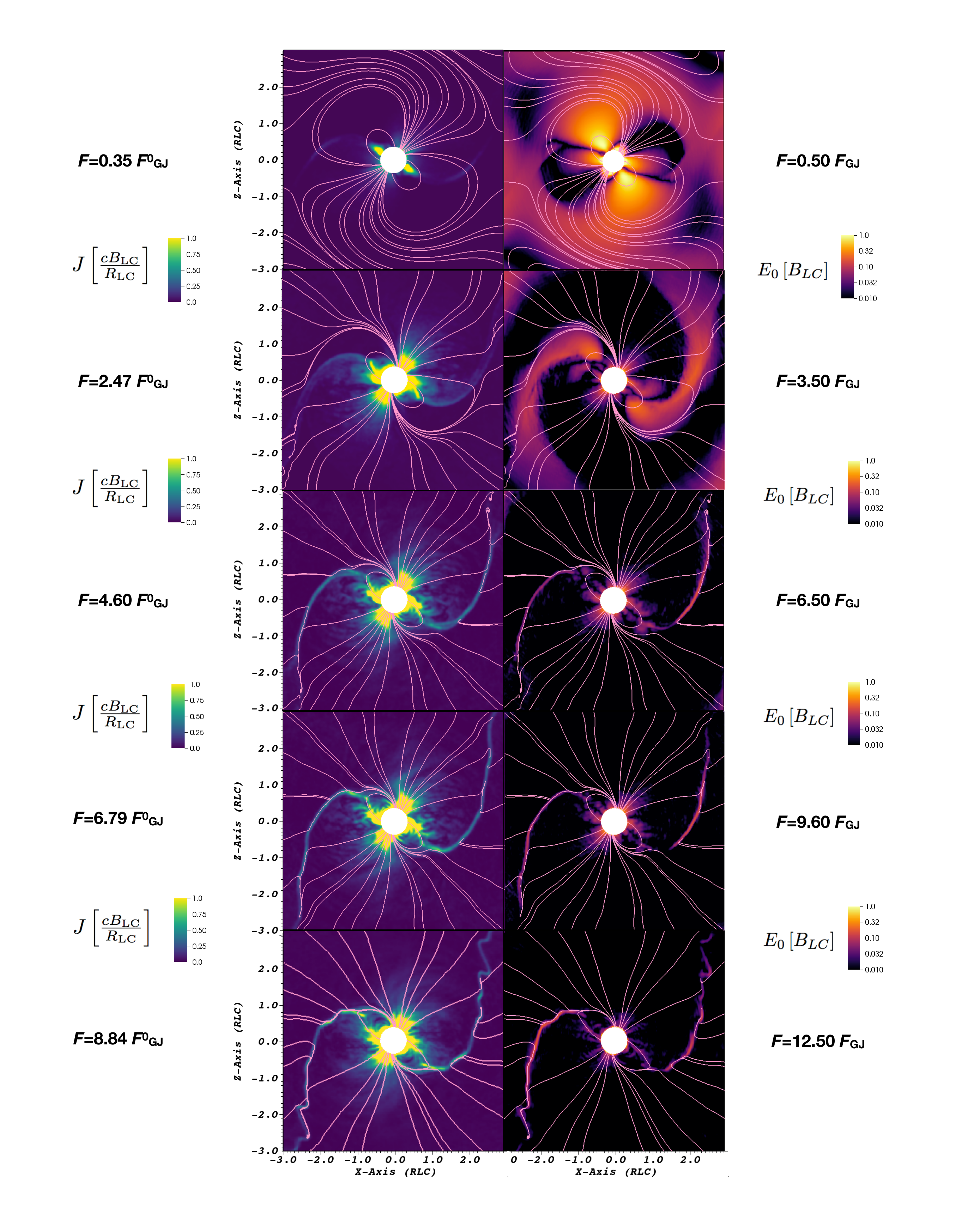}
  %\vspace{-0.2in}
  \vspace{-0.1in}
  \caption{This Figure shows how $J$ and $E_0$ (this last one defined in Equations \ref{eq:eknot1} and \ref{eq:eknot2}) varies with the injection rate $\mathcal{F}$ only from the neutron star surface. The field lines in the background are the magnetic field lines. The gradual screening of $E_0$ and the formation of the force-free current structure are shown.}
  \label{fig:screening}
\end{figure*}

%%%%%%%%%%%%%%%%%%%%%%%%%%%%%%%%%%%%%%%%%%%%%%%%%%%%%%%%%%%%%%%%%%
Studying how the magnetosphere changes its configuration while the number of particles injected increases is interesting because it shows possible configurations in which a pulsar can operate. However, exact solutions reflecting what happens in a real pulsar can be obtained only having a detailed knowledge of the pair production processes in the magnetosphere, works like \cite{Tim13} go in this direction. In our study we can observe how the magnetosphere behaves if the particles are injected only at the surface, but without capturing the specific physical mechanism behind the injection. \newline
The magnetic field structure, shown in Figure \ref{fig:screening}, begins to resemble the force-free structure already from $\mathcal{F}\sim 5\mathcal{F}_{\rm GJ}$, while for  $\mathcal{F}=0.5\mathcal{F}_{\rm GJ}$ it is very close to the vacuum one.
The evolution of the current configuration seems to follow the magnetic field structure, in particular once the magnetic field resembles the force-free one, a clear separatrix/Y-point/current sheet configuration is present.
As accelerating electric field we consider $E_0$ (\citealt{Gruz08, Li12}) that is defined as:
\begin{equation}\label{eq:eknot1}
B_0^2-E_0^2=\mathbf{B}^2-\mathbf{E}^2
\end{equation}
\begin{equation}\label{eq:eknot2}
B_0E_0=\mathbf{B}\cdot \mathbf{E}
\end{equation}
with $E_0\geq 0$. The electric field gets gradually screened with higher particle injection, but we can see that the regions that are hardest to screen are the polar cap outflow region and the separatrix region. When the current sheet region is formed, it never gets completely screened.

%%%%%%%%%%%%%%%%%%%%%%%%%%%%%%%%%%%%%%%%%%%%%%%%%%%%%%%%%%%%%%%%%%
\section{Comparison between the solutions approaching the force-free limit}\label{sec:ff_solutions}
%%%%%%%%%%%%%%%%%%%%%%%%%%%%%%%%%%%%%%%%%%%%%%%%%%%%%%%%%%%%%%%%%%
In this section, we present the solution approaching the force-free limit ($\mathcal{F}=12.5\mathcal{F_{\rm GJ}}$) obtained with particles supplied only at the surface for $\alpha=45^\circ$, as described in section \ref{sec:energy}.
%%%%%%%%%%%%%%%%%%%%%%%%%%%%%%%%%%%%%%%%%%%%%%%%%%%%%%%%%%%%%%%%%%
\subsection{The macroscopic quantities}\label{sec:macroscopic}
\begin{figure*}
  \singlespace
  \centering
  %\hspace{-.5in}
  \includegraphics[width=1.\textwidth]{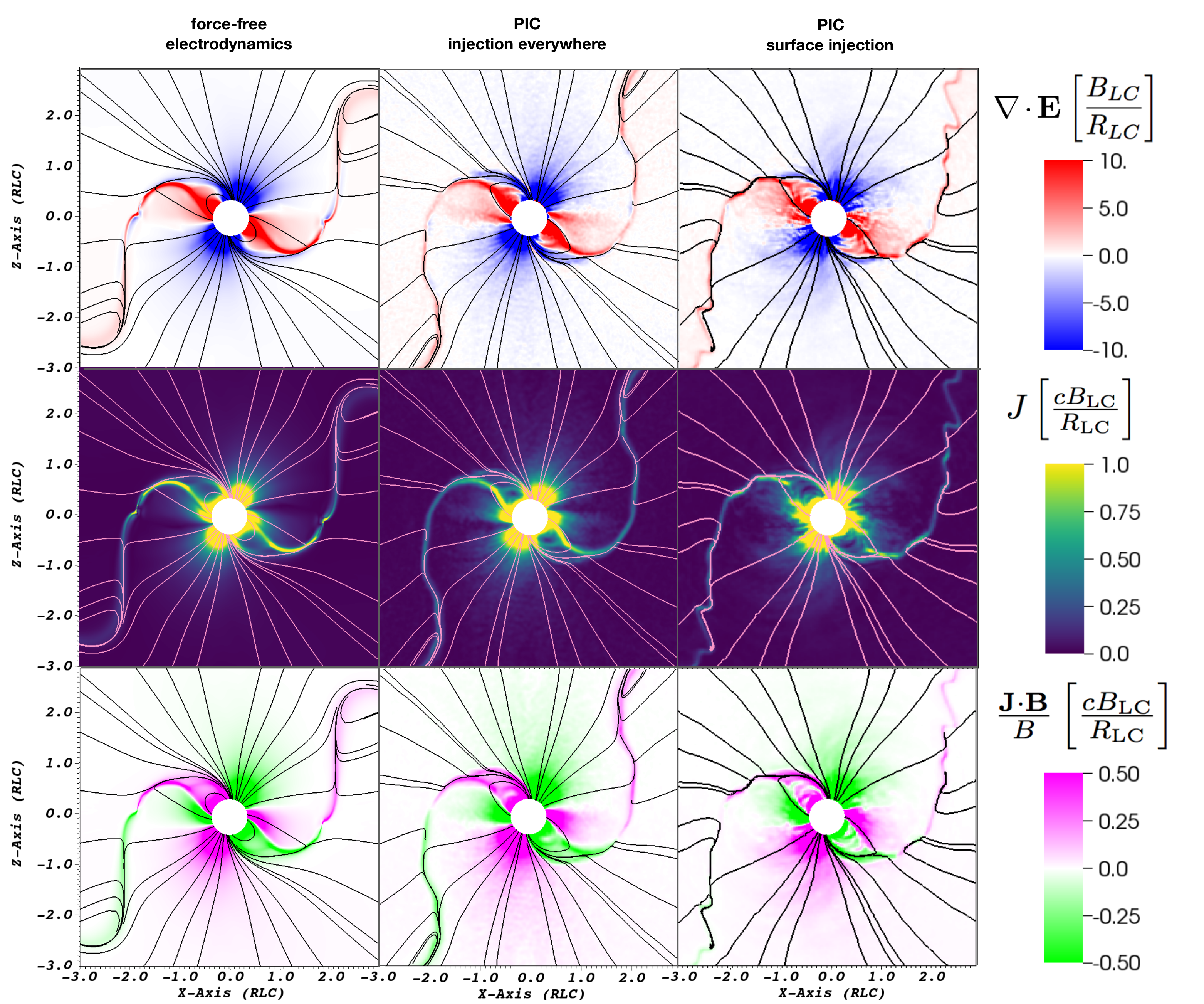}
  %\vspace{-0.2in}
 \vspace{-0.1in}
  \caption{The divergence of the electric field, the absolute value of the current density and the projection of the current density on the magnetic field lines for magnetospheres close to the force-free limit obtained with three different kinds of techniques: force-free electrodynamics, PIC with particles injected everywhere in the domain, PIC with particles supplied only at the surface. As it can be seen, these plots are very similar.}
  \label{fig:forcefree}
\end{figure*}

\begin{figure*}
  \singlespace
  \centering
  \includegraphics[width=1.\textwidth]{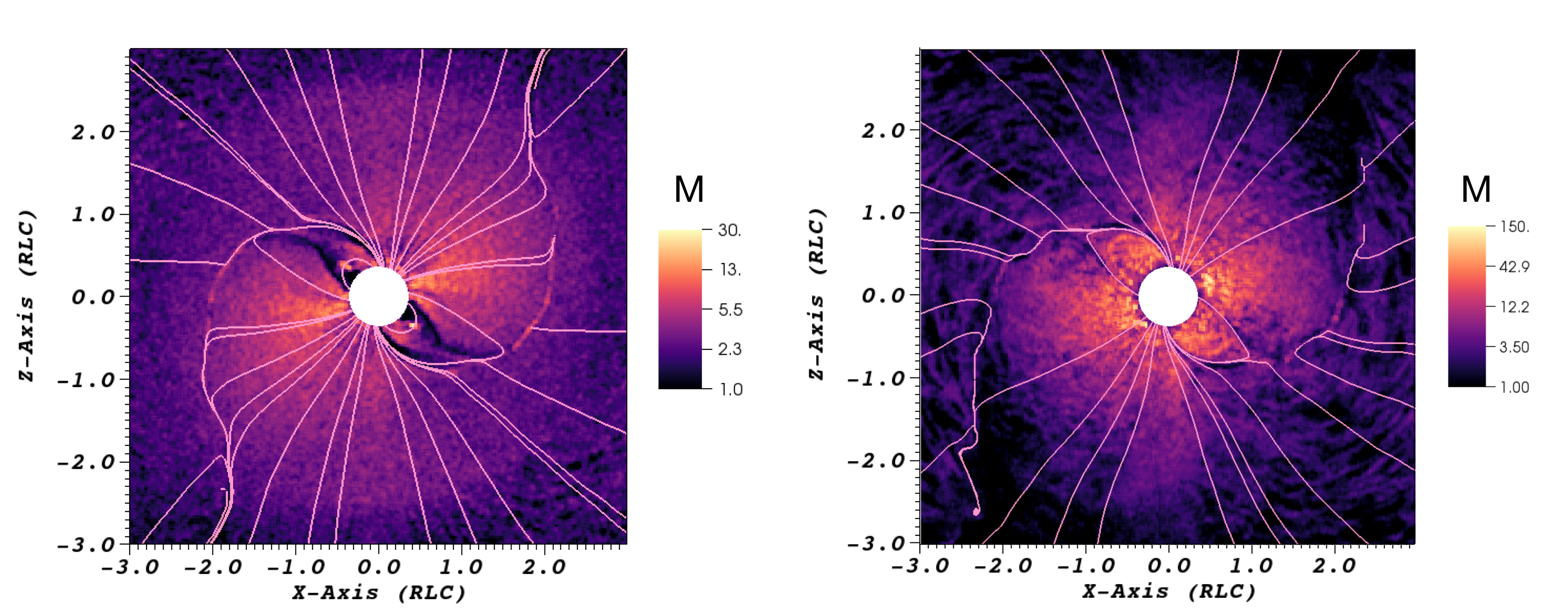}

 \vspace{-0.1in}
  \caption{Multiplicity profile for PIC simulations close to the force-free limit: on the right particles are supplied only at the star's surface, on the left particles are injected everywhere. When particles are injected only at the surface the multiplicity is higher out to a radius of $1R_{LC}$, similar up to $2R_{LC}$ and lower up to $3R_LC$. The color scale is logarithmic.}
  \label{fig:multiplicity}
\end{figure*}

%%%%%%%%%%%%%%%%%%%%%%%%%%%%%%%%%%%%%%%%%%%%%%%%%%%%%%%%%%%%%%%%%%
In Figure \ref{fig:forcefree} we compare the solution obtained with force-free electrodynamics to two PIC simulations approaching the force-free limit: one injecting particles everywhere in the simulation domain and one injecting particles only close to the star surface. The $\nabla\cdot \mathbf{E}$ (which represents the charge density), the total current and its sign are very similar in all the solutions. 
The situation changes when we look in detail at how these structures are sustained.
The two solutions are obtained with a different $\mathcal{F}$. This is due to the different way in which the injected particles are distributed in the two injection schemes. For injection everywhere, particles are placed at larger radii, filling a greater volume than the injection from the surface configuration. The more the particles are injected at larger distance, the less they contribute to the force-freeness of the inner magnetosphere because many quickly leave the domain. However, the injection scheme is not crucial to obtain a force-free configuration, as we can see from the very similar values of charge and current densities.
We define the multiplicity $M$ as the number of particles present per charge at a given location. 
\begin{equation}\label{eq:multiplicity}
M = \frac{N_{ele}+N_{pos}}{| N_{ele}-N_{pos} |}
\end{equation}  
When the particles are injected at the surface, the multiplicity reaches higher values close to the star, but lower values at larger radii. This can be seen in Figure \ref{fig:multiplicity}.

\begin{figure*}
  \singlespace
  \centering
  %\hspace{-.5in}
  \includegraphics[width=1.\textwidth]{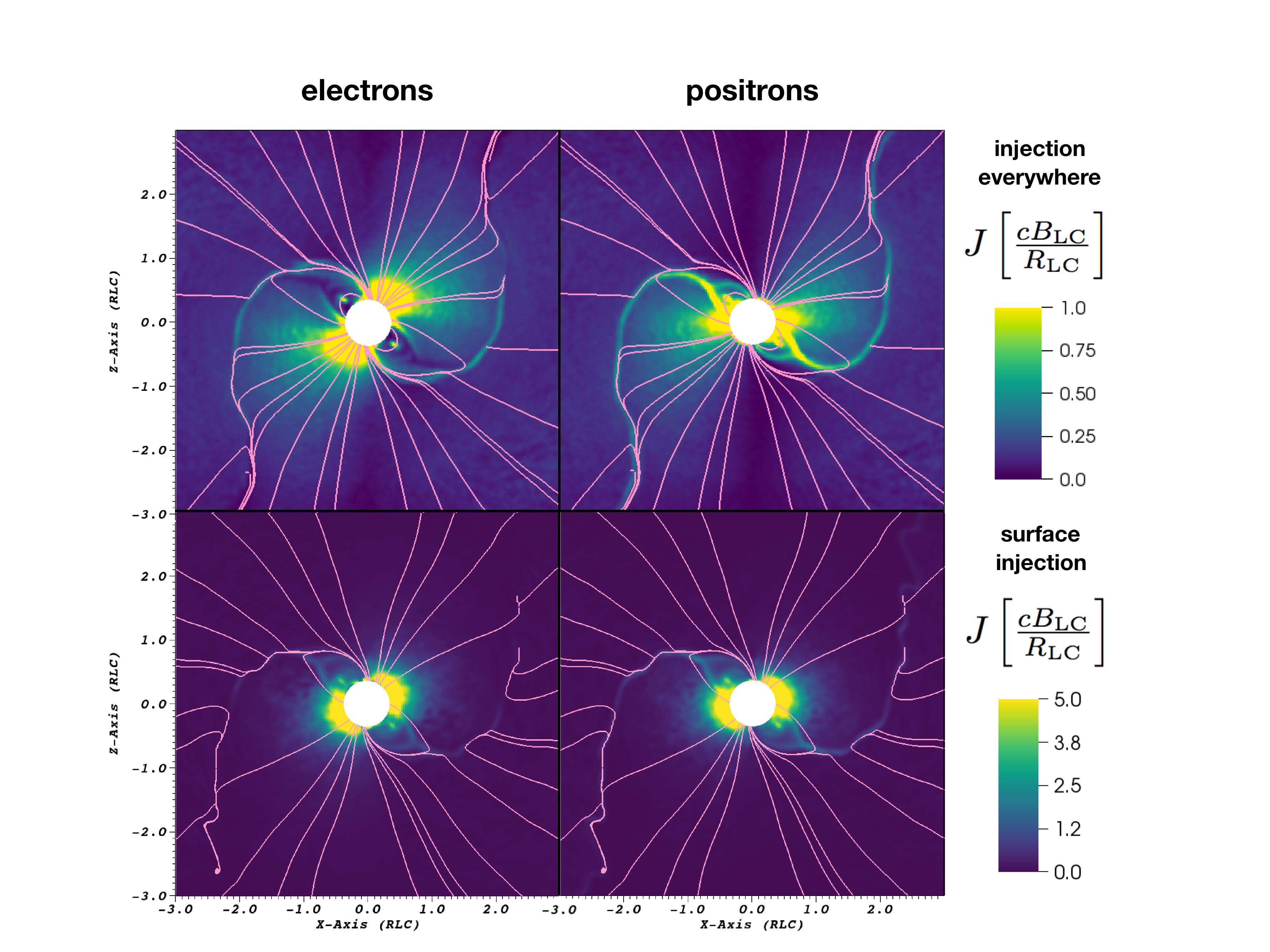}
  %\vspace{-0.2in}
 \vspace{-0.1in}
  \caption{Electronic and positronic modulus of the current densities for PIC simulations close to the force-free limit: one is with particles supplied only at the star's surface, the other is with particles injected everywhere. Notice that the scale values are different: in the case of injection from the surface the two components are much more intense.}
 \label{fig:abs-cur}
\end{figure*}

As we described in section \ref{sec:methods}, our code accounts for the different current contributions of electrons and positrons. We plot their absolute value components in Figure \ref{fig:abs-cur}. 

\begin{figure*}
  \singlespace
  \centering
  %\hspace{-.5in}
  \includegraphics[width=1.\textwidth]{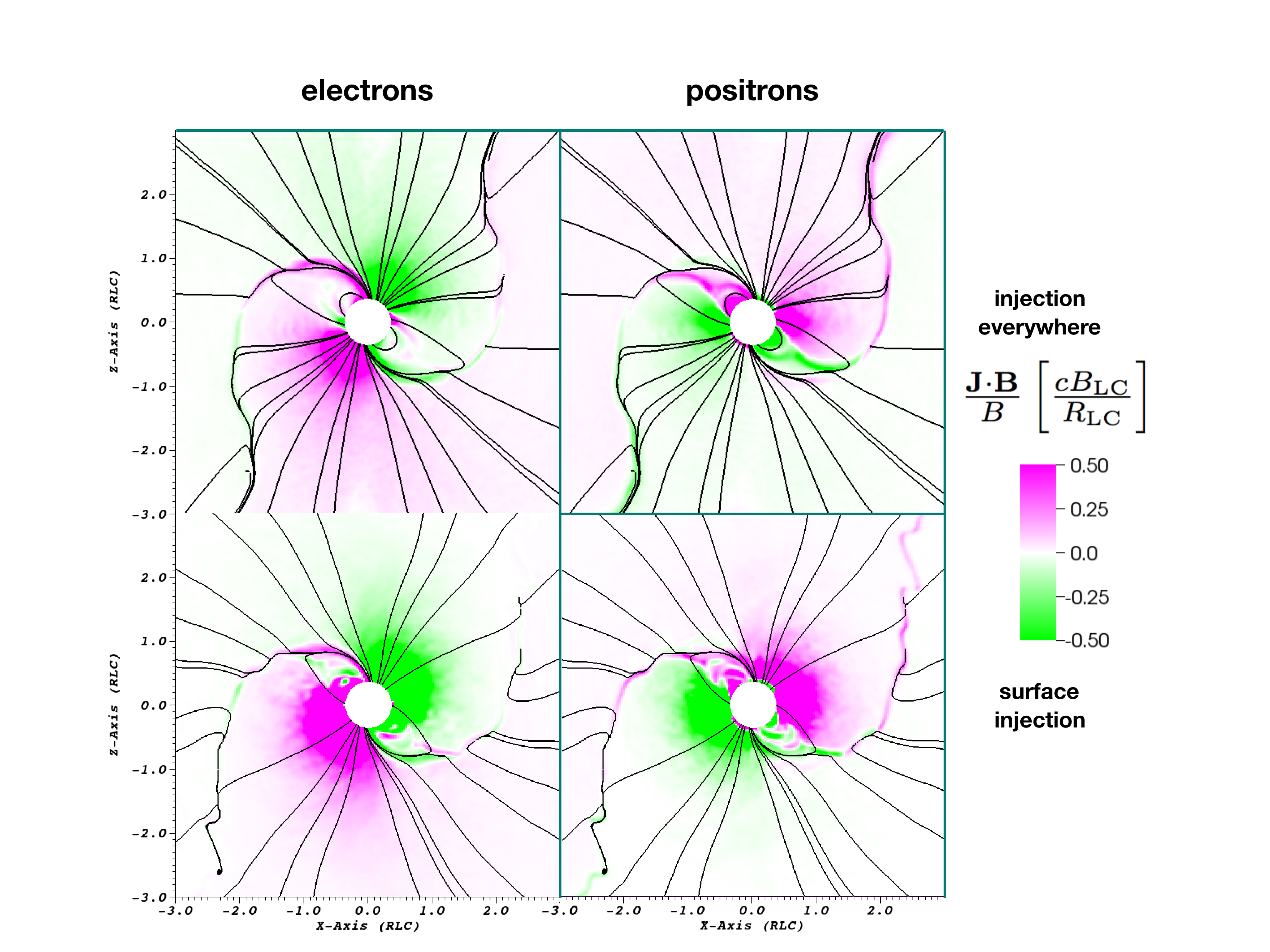}
  %\vspace{-0.2in}
 \vspace{-0.1in}
  \caption{Electronic and positronic projection of the current density on the magnetic field lines for PIC simulations close to the force-free limit: one is with particles supplied only at the star's surface, the other is with particles injected everywhere.}
  \label{fig:jpol-comp}
\end{figure*}

In the case when particles are injected everywhere, electron currents are present mostly in the negatively charged regions, while positron currents are present mostly in the positively charged regions (for the charge density plot see Figure \ref{fig:forcefree}). When we inject particles from the surface, we notice that the electrons (positrons) have an important current contribution even in positively (negatively) charged regions (Figure \ref{fig:abs-cur},\ref{fig:jpol-comp}). As we saw in Figure \ref{fig:forcefree}, the total current densities are very similar, but the difference in electron and positron current densities indicates that when we inject particles only from the surface there are regions where electrons and positrons are streaming in the same direction with electron and positron currents almost cancelling each other.
This is possible if these particles are injected in a zone where the accelerating electric field is screened enough to not reverse their initial velocities acquired by the numerical heating. The same qualitative behavior is present in the electron-positron pair cascades at the polar cap. In fact, in polar cap cascades most of the pairs are produced above the pair formation front wth some initial Lorentz factors \citep{HarMus01,TimHar15}. When particles are injected everywhere in the domain, they are supplied wherever they are needed. Instead when particles are injected only from the surface, they need to arrange themselves in a different way to satisfy the current and charge density requirements of the magnetosphere. \newline
In Figure \ref{fig:jpol-comp} where we plot $\mathbf{J} \cdot \mathbf{B}/B$ for the electron and positron components, this scenario becomes clear. 
For the simulation with injection from the surface, we can see that $\mathbf{J} \cdot \mathbf{B}/B$ indicates counter streaming flows (where the two components have the same color in the same region) only on the negative branch of the separatrix and in a thin layer just above the neutron star surface, where particles are injected.
In the same simulation, there is also a clear component of positrons flowing out from the polar cap together with the electrons that is very weak in the simulation with the particles injected everywhere. In the electron component the positive branch of the separatrix that connects the star surface to the Y-point changes sign: in the surface injection case, on this branch the electrons are flowing outward from the star, while in the injection everywhere case they are flowing in. Therefore, we see that the zones with availability of pairs (where the pair creation happens) greatly influence the underlying currents of the single species and this has important consequences that we will outline in the Discussion (Section \ref{sec:discussion}).

%%%%%%%%%%%%%%%%%%%%%%%%%%%%%%%%%%%%%%%%%%%%%%%%%%%%%%%%%%%%%%%%%%
\subsection{The particle trajectories}\label{sec:trajectories}
\begin{figure*}
  \singlespace
  \centering
  %\hspace{-.5in}
  \includegraphics[width=0.8\textwidth]{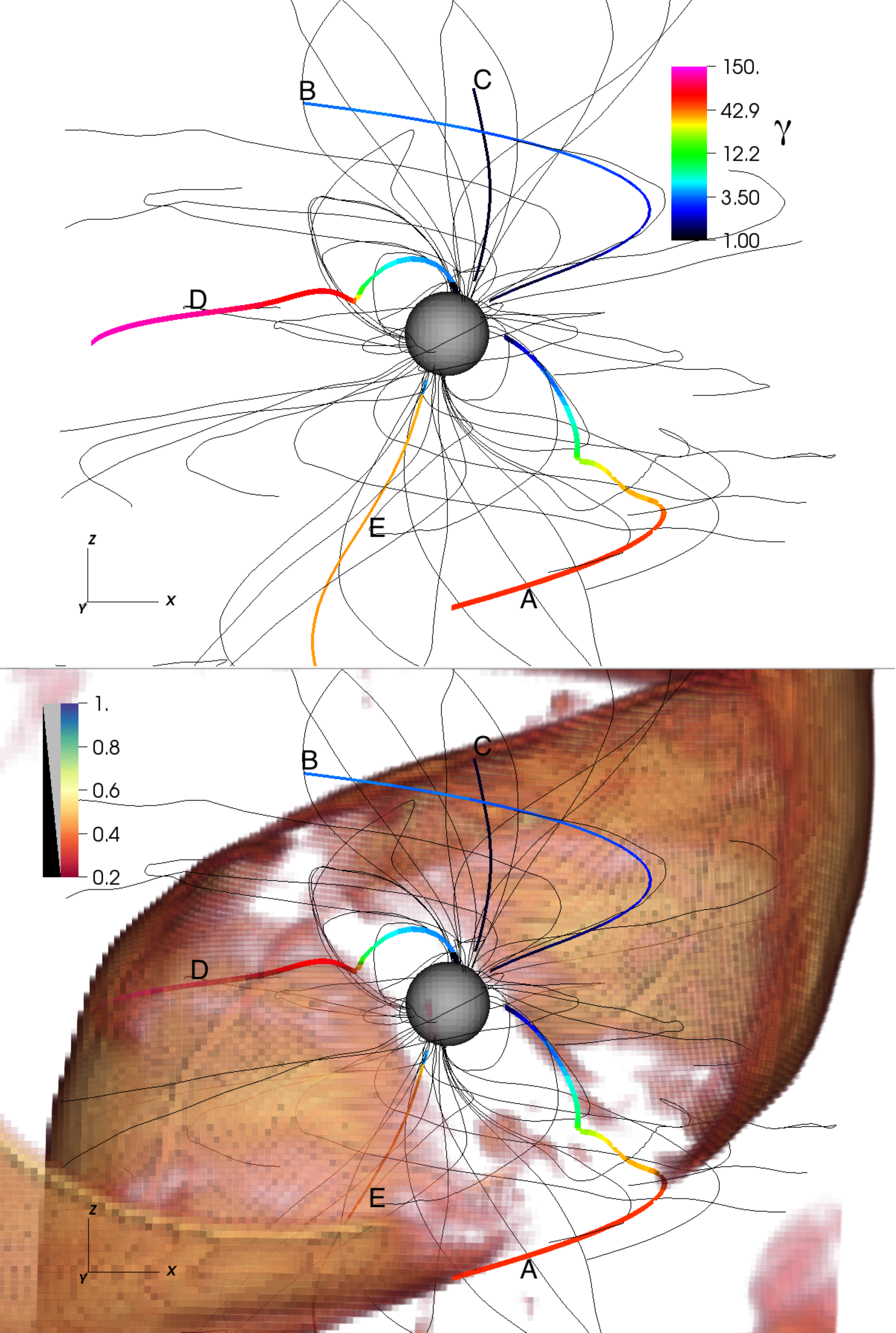}
  %\vspace{-0.2in}
 %\vspace{-0.1in}
  \caption{Most common particle trajectories in the simulation approaching force-free with particles supplied close to the surface. The color on the trajectories represents the Lorentz factor ($\gamma$). A) is a ``not so highly" accelerated positron. B) is a positron flowing out from the polar cap flow at low energy. C) is an electron flowing out from the polar cap flow at low energy. D) is a ``highly accelerated" positron. E) is an intermediate energy electron flowing out from the polar cap. In the picture on the bottom, we have a volume rendering of $E_0$ (Equation \ref{eq:eknot1}, \ref{eq:eknot2}) that identifies the current sheet: we can see that the difference in acceleration between A) and D) is due to the strength of $E_0$ on the trajectory. This non uniformity in $E_0$ is found only through PIC simulations and it can be useful to model the $\gamma$-ray emission. All the trajectories are in the corotating frame.}
  \label{fig:std_traj}
\end{figure*}

\begin{figure}
  \singlespace
  %\centering
  %\hspace{-.5in}
  \includegraphics[width=0.5\textwidth]{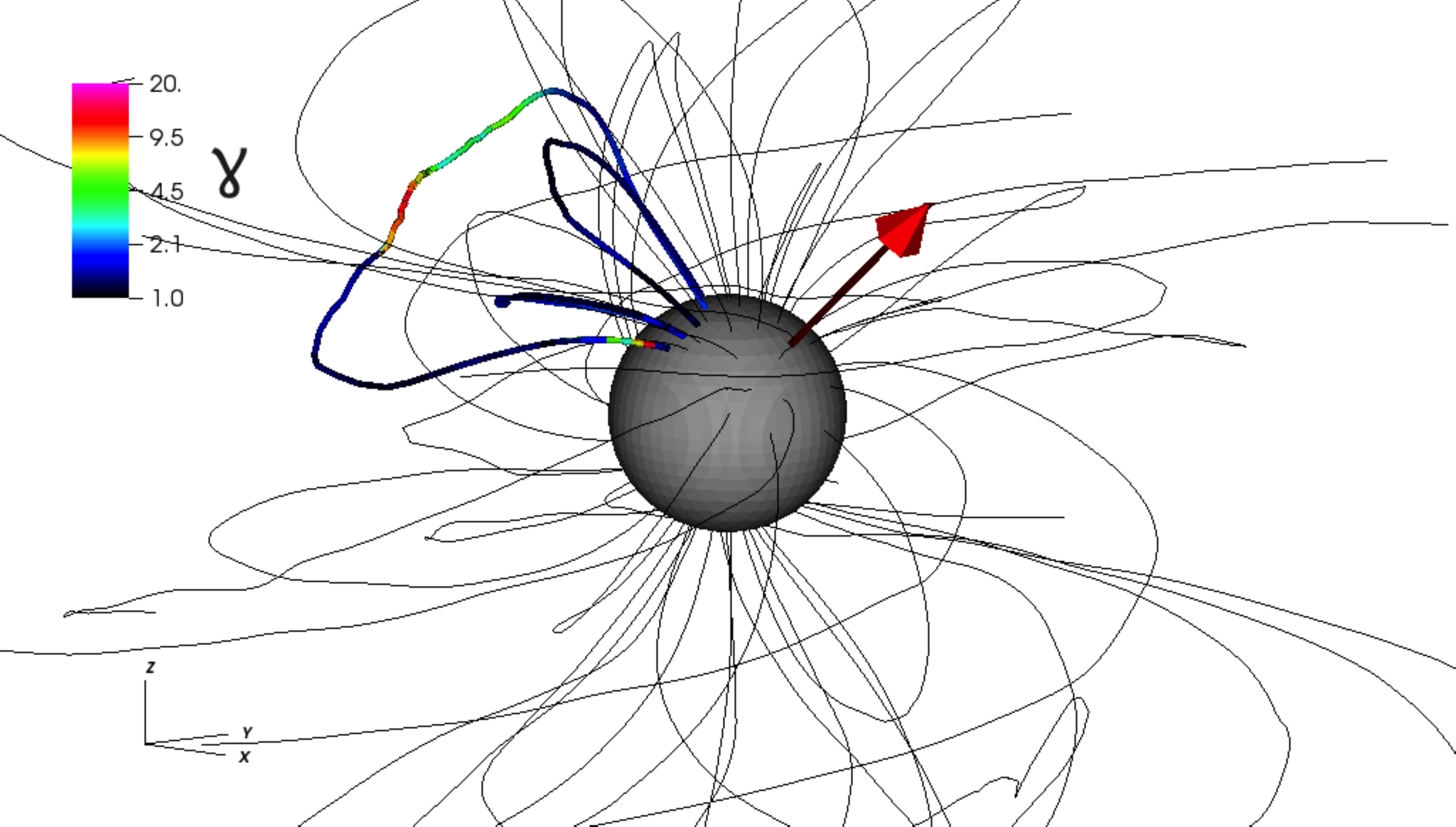}
  %\vspace{-0.2in}
  \caption{Electron trajectories falling back on the star from different heights. The Lorentz factor $\gamma$ is the color on the trajectory. We added a red arrow representing the magnetic moment, because the zoomed region could be difficult to identify. Note that the color scale has a different range with respect to all the others shown in this paper for trajectories. This is because of the low energy of these particles. The trajectories are in the corotating frame.}
  \label{fig:falling}
\end{figure}

\begin{figure}
  \singlespace
  %\centering
  %\hspace{-.5in}
  \includegraphics[width=0.5\textwidth]{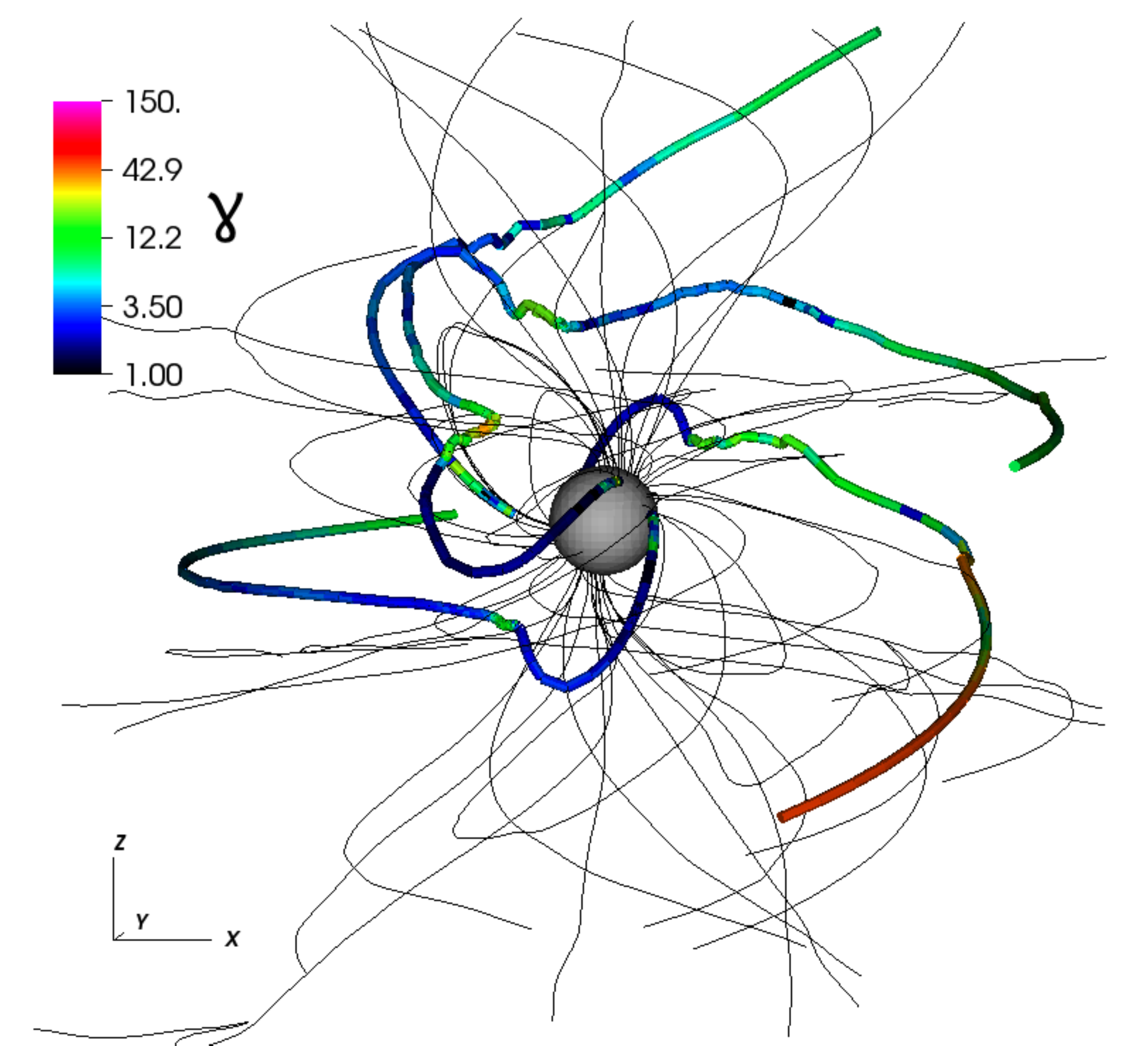}
  %\vspace{-0.2in}
  \caption{Electron trajectories circling around the Y-point and the light cylinder. The Lorentz factor $\gamma$ is the color on the trajectory. The trajectories are in the corotating frame.}
  \label{fig:circling}
\end{figure}

\begin{figure*}
  \singlespace
  \centering
  %\hspace{-.5in}
  \includegraphics[width=1.0\textwidth]{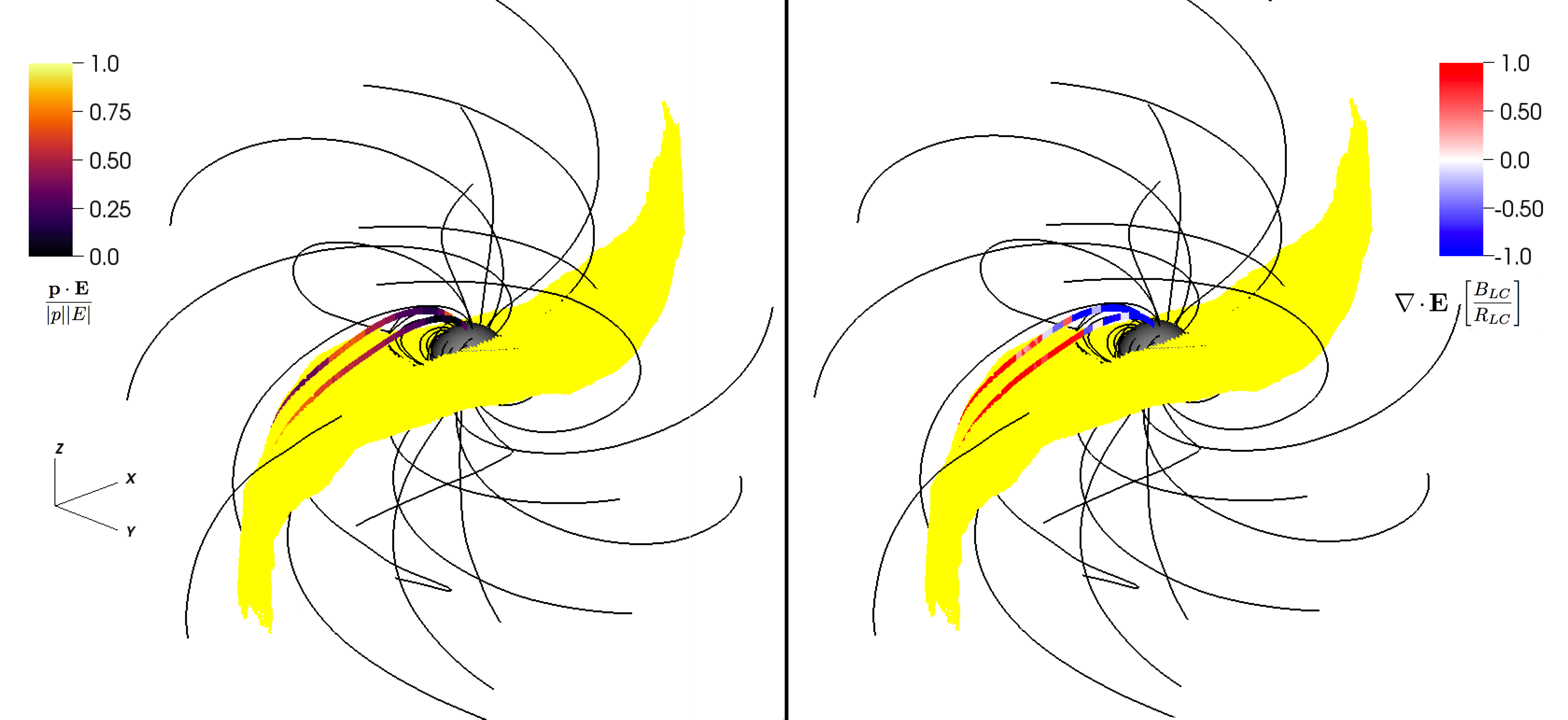}
  %\vspace{-0.2in}
 %\vspace{-0.1in}
  \caption{Positron trajectories flowing from the polar cap into the current sheet. The star with the magnetic field lines is plotted on the background. The yellow opaque surface is the pulsar current sheet. We used an opaque profile instead of a volume rendering profile (as in Figure \ref{fig:std_traj}) in order to facilitate seeing that the trajectories shown are initially outside the current sheet. We selected the current sheet according to Equation \ref{eq:eknot-select}. On the left panel the color on the trajectories is the normalized projection of $\mathbf{p}$ on $\mathbf{E}$, as defined in Equation \ref{eq:pdotE}. We can see that the value is larger than 0.5 in many parts of the trajectories. On the right panel the color is the divergence of the electric field along the trajectories. We can see that these positrons are transitioning from a negatively charged region, to a positively charged region. The trajectories are in the corotating frame.}
  \label{fig:pos}
\end{figure*}

\begin{figure*}
  \singlespace
  \centering
  %\hspace{-.5in}
  \includegraphics[width=\textwidth]{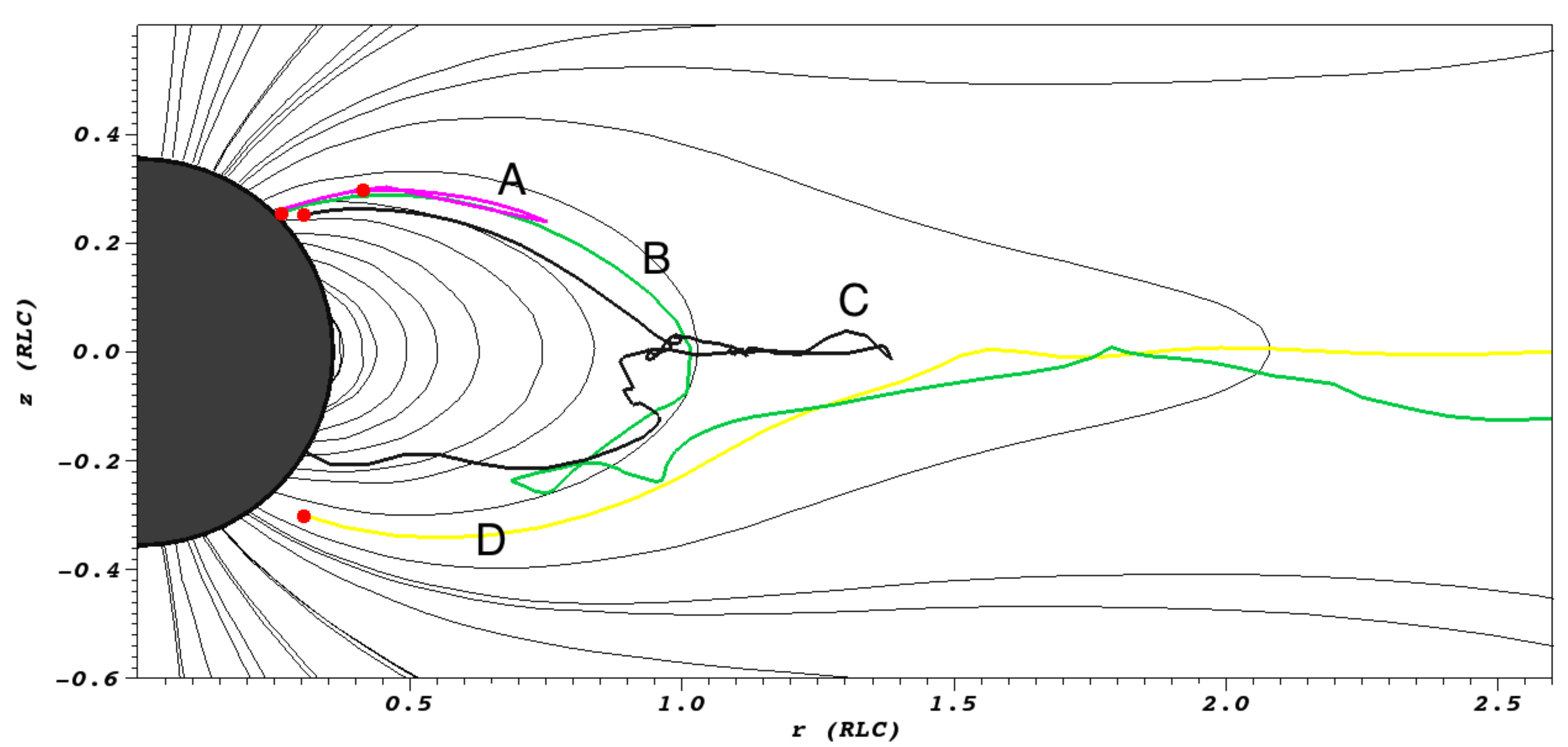}
  %\vspace{-0.2in}
  \caption{The particle trajectories that we presented above, but for the aligned rotator. For clarity, we removed the azimuthal component of their trajectory. The magnitude of the azimuthal components were similar to the trajectories shown in the Figures \ref{fig:falling}, \ref{fig:circling}, \ref{fig:pos}, for the $45^\circ$ case. The red dot indicates where the particle is injected. All these trajectories had large azimuthal components. The color is a label to help distinguish their intricate trajectories. A) is one of the electrons that starts from the polar cap outflow and get turned back into the separatrix and the returning current. B) is one of the electrons that reaches the Y-point, circles for a while and then flies out following another field line. C) is one of the electrons that reaches the Y-point, circles with huge azimuthal components, and then falls back to the star losing energy. D) is one of the positrons that starts in the polar cap flow, close to the separatrix, crosses field lines and then enters the current sheet. }
  \label{fig:aligned}
\end{figure*}

\begin{figure*}
  \singlespace
  \centering
  %\hspace{-.5in}
  \includegraphics[width=\textwidth]{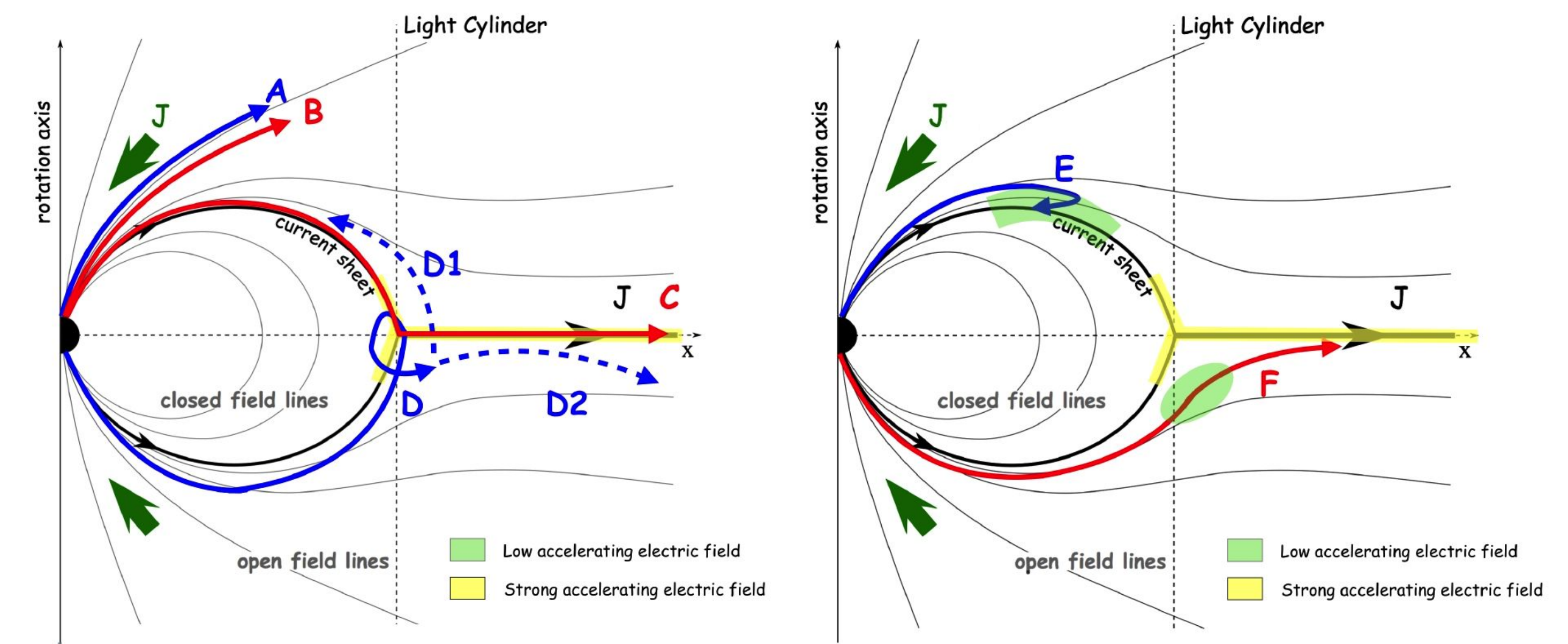}
  %\vspace{-0.2in}
  \caption{In the sketch above we illustrate for the aligned rotators case the electron and positron trajectories summarized at the end of Section \ref{sec:trajectories}. The action of the low electric fields acting on low energy particles is highlighted by the green quasi transparent shapes. The stronger electric fields are highlighted by the yellow quasi transparent shapes. A) an electron trajectory that flows out from the polar cap. B) one of the positron trajectories that flow together with the electrons out of the polar cap. C) trajectory of a positron accelerated at high energy by the electric field at the Y-point and in the reconnecting current sheet. D) a trajectory of an electron that arrives close to the Y point and is bounced back by the same electric field that accelerates the positrons at very high energy. This bouncing makes the electron circle all around the $R_{\rm LC}$ and gain energy. Then the electron can either fall back to the star D1), or fly away D2) depending on where this circling around the $R_{\rm LC}$ takes it. E) a trajectory of a low energy electron that crosses magnetic field lines towards the return current region. Eventually, this electron is deflected by a low electric field and returns to the star. Electrons approximately on the same trajectory but with a higher energy are not deflected. F) a trajectory of a positron that flows out of the polar cap at low energy as B), but outside the $R_{\rm LC}$ it crosses the field lines and enters in the current sheet far from the Y-point. The figure is a modified version of a figure from \cite{Tim2006}.}
  \label{fig:sketch}
\end{figure*}

%%%%%%%%%%%%%%%%%%%%%%%%%%%%%%%%%%%%%%%%%%%%%%%%%%%%%%%%%%%%%%%%%%
We studied the trajectories of the particles in our simulation approaching the force-free limit with injection from the surface ($\mathcal{F}=12.5\mathcal{F_{\rm GJ}}$). First we describe trajectories followed by the majority of the particles. In general the most energetic particles (the ones that reach $\gamma$ from $\sim 50$ up to $\sim 180$ in the $45^\circ$ close to force-free simulation) are mainly positrons accelerated along the field lines that constitute the separatrix\textbackslash Y-point\textbackslash current sheet complex. The particles gain most of their energy in the proximity of the Y-point. At intermediate energies ($\gamma\sim40$) we find the electrons flowing from the polar cap. Then at low energies ($\gamma<30$) we find the bulk of the flow with electrons and positrons generally flowing out together. As expected, positrons are dominant in positively charged regions and electrons are dominant in negatively charged regions. In Figure \ref{fig:std_traj}, we see some examples of these trajectories. All the trajectories are shown in the corotating frame. 
\newline
Studying PIC simulations of the pulsar magnetosphere is interesting because it can provide solutions to problems that are present in the force-free electrodynamics limit. One of these problems, is how the current structure of the force-free configuration could be sustained by particles in real pulsars, outside the strict force-free limit. It is reasonable to assume that the field structure of a pulsar magnetosphere is stationary, therefore, the amount of charge in the magnetosphere should remain constant. Because of the charge conservation it follows that the current leaving the star should be balanced by a current entering the star. When \cite{CKF1999} found the first force-free solution for a dipolar magnetic field, the currents were going from the star to infinity and coming from infinity to the star, one through the polar cap flow, the other mainly through the current sheet and separatrix, and a smaller part on a few open magnetic field lines close to the last open magnetic field lines. The surface charge density of the current sheet has some puzzling features. The charge of an aligned force-free magnetosphere at the Y-point should be negative inside the Y-point and positive outside of it \citep{Lyu90,Tim2006}. However, the current is continuous through the Y-point, but its composition should change to obtain a charge of a different sign. It is not clear how electrons can flow back to the star and positrons flow into the current sheet both from the Y-point, especially when particles are injected only at the surface. An outer gap \citep{CRS76} was thought to provide electrons flowing backward and positrons flowing outward where the charge density changes sign. Studying the particle trajectories, we have found that the pulsar magnetosphere does not need pair production in an outer gap to fill the magnetosphere and maintain the charge density distribution of the separatrix/Y-point/current sheet complex. 
As we said above, there is an outward flow of electrons from the polar cap. Some of them have $\gamma \sim 1$ and flow very close to the region where the current changes sign. There a low electric field drags part of the low energy electron distribution into the returning current and separatrix region, where the majority of them form the returning current by the electric field that reverse their velocity (Figure \ref{fig:falling}). The particles with a higher energy are not affected by this because this accelerating electric field is too small.
 This phenomenon does not happen at a specific height, but it appears continuous up to the Y-point. The crossing of field lines happens in a region where $\lambda_{sd}$ is not resolved, therefore this noise could in principle affect the trajectories. We checked a randomly selected sample of all the electron trajectories starting from the region where the electrons that turn back originate. We found that $\sim70\%$ of the electrons coming from this region are turned back inside the light cylinder. This behavior is different from the action expected of kicks due to random noise; thus we believe that this phenomenon results from non-fluctuating, low electric fields.\newline
Another phenomenon concerns the outgoing electrons.  Some electrons remain stuck at the Y-point and they circle all around the light cylinder, see Figure \ref{fig:circling} (a similar behavior was shown also in \cite{Cerutti2016}). This happens because the solution tends toward the force-free one where the sign of the charge density is negative before the Y-point and positive after it. The resulting electric field accelerates positrons and deflects the electrons. In this motion the electrons get energized, $40 \lesssim \gamma \lesssim 90$. 
Once they are in this regime electrons have two possibilities: either falling back toward the star (and they mix with the electrons of Figure \ref{fig:falling}) or flying out following other field lines (they do not usually fly far out into the current sheet). When they fall back they lose their energy by radiation reaction (the accelerating fields in that region are not strong enough to sustain the Lorentz factor they had reached). Beyond the $1.5R_{\rm LC}$ there are very few electrons that turn back and the number of these returning ones decrease drastically with distance, in contrast with what was presented in \cite{Cerutti2015}. However, the \cite{Cerutti2015} simulation setup is different from ours; for example, they do not include radiation reaction.\\
To complete the picture, we must understand the origin of the positrons that support the charge density change of sign through the Y-point. Most of the positrons in the current sheet come along the separatrix; but extra positrons are needed in the current sheet to account for the current of the returning electrons inside the Y-point. They come from the polar cap flow (they are flying out with the electrons), close to the returning current region and the separatrix and they cross field lines outside the light cylinder to enter into the positively charged region and then the current sheet (Figure \ref{fig:pos}). We checked that positrons indeed cross magnetic field lines (for the returning electrons it was obvious because of the shapes of their trajectories) looking at the cosine of the angle between the particle momentum and the local electric field outside of the current sheet
\begin{equation}\label{eq:pdotE}
 \frac{\mathbf{p}\cdot \mathbf{E}}{pE}           
 \end{equation}
with $\mathbf{p}$ particle momentum and $\mathbf{E}$ electric field.
We use this criterion because in the force-free limit, where particles flow exactly along the magnetic field lines and $\mathbf{E}\bot\mathbf{B}$, $\mathbf{p}$ has components along $\mathbf{B}$ and $\mathbf{E}\times\mathbf{B}$. $\mathbf{E}\times\mathbf{B}$ keeps the particle on the rotating field line. In this limit, the Expression \ref{eq:pdotE} is always 0. 
In Figure \ref{fig:pos}, we see that the expression \ref{eq:pdotE} becomes singificantly $>0$. This happens in regions where $\mathbf{E}\cdot\mathbf{B}\ll BE$ (where $\mathbf{E}$ is mostly perpendicular to $\mathbf{B}$).
In fact, we identified in our PIC simulations the regions where $\mathbf{E}$ has significant components parallel to $\mathbf{B}$ (the yellow opaque volume in Figure \ref{fig:pos}) using
 \begin{equation}\label{eq:eknot-select}
 \frac{\mathbf{E}\cdot \mathbf{B}}{BE}>0.15           
 \end{equation}
and the field line crossing happens outside of this region. The region defined by the Expression \ref{eq:eknot-select} traces quite well the regions of reconnecting $\mathbf{B}$. 
The last three kinds of trajectories that we just described involve the crossing of magnetic field lines. The theoretical gyro radius of these low energy particles is very small while the use of strong radiation reaction forces makes it even smaller. Therefore, the corresponding crossing of the magnetic field lines is not driven by a large gyro radius due to the use of a low magnetic field, but by unscreened electric fields.
%The gyro radius of the bulk of the particles is very small and the strong radiation reaction is implemented, therefore the crossing of the magnetic field lines is due to non ideal electric fields, especially if the particles have low kinetic energy.}
\newline
So far, we showed trajectories for $\alpha = 45^\circ$ case. However, the most well studied case is the aligned rotator and we looked for the same trajectories in this case as well and we show them in Figure \ref{fig:aligned}. In this case, the electrons circling around the Y-point form a cloud of negative charge that appears as an increase of the negative charge density where the separatrix touches the Y-point. This behavior was first noted in force-free electrodynamics simulations by \cite{Tim2006}.      
Therefore, we can safely say that this current structure is sustained by particles crossing field lines mainly inside $1.0 R_{LC}$ and for sure inside $2.0 R_{LC}$. The main mechanisms are all driven by low electric fields operating on the low energy part of the particle distribution. 
For the nearly orthogonal rotator case, the structure of the current sheet is very different \citep{Kala12}, therefore we decided to reserve this study for the future. Summarizing, we can say that the pulsar magnetosphere structure approaching the force-free solution with particles injected from the surface has these features:\\
1) Electrons and positrons stream outward together in the polar cap outflow.\\
2) The electrons that flow back to the star cross field lines, either from the polar cap outflow into the returning current region inside the light cylinder or after circling around the Y-point. \\
3) Positrons flow out on the separatrix and get accelerated close to the Y-point into the current sheet. Some positrons enter in the current sheet beyond the Y-point crossing field lines.\\
In Figure \ref{fig:sketch} we show a sketch for an aligned rotator to help the reader understand the particle trajectories outlined above. The action of the non ideal electric field is indicated.
 
 %%%%%%%%%%%%%%%%%%%%%%%%%%%%%%%%%%%%%%%%%%%%%%%%%%%%%%%%%%%%%%%%%%
\section{Discussion}\label{sec:discussion}
In this section we discuss the results we obtained in the context of pulsar magnetosphere theory. First of all, we should note again the limitations of PIC codes in capturing the physical quantities of the pulsar magnetosphere. The open field voltage from Equation \ref{eq:openvolt} for a pulsar with $B_0\sim10^{12}G$ and $P\sim0.1s$ would be $\gamma_{max}\sim10^9$, while we and all the previous studies with PIC simulations (e.g \citealt{Chen2014}; \citealt{Belyaev2015b}; \citealt{Cerutti2016}; \citealt{Philippov17}; \citealt{Kalapotharakos2017b}) use $100\lesssim\gamma_{max}\lesssim1000$. This is necessary because we need to resolve $\omega_p$ everywhere in our system, as we explained in Section \ref{sec:methods}. Therefore, the particle energy distribution is squeezed into a narrow range and it cannot be simply linearly stretched or shifted to higher energies\footnote{In \cite{Kalapotharakos2017b}, we try to recover the actual high energy end of the particle distribution.}. \newline
Some of the previously referenced works on PIC pulsar magnetospheres how their results depend on the number of particles injected. We showed that many properties and accelerating gaps are connected to the number of particles injected into the magnetosphere, and for high injection rate the magnetosphere reached the force-free limit for the whole range of inclination angles ($0^{\circ}$, $45^{\circ}$, $85^{\circ}$). In our case, we are confident in claiming that the current composition we discussed is characteristic of a magnetosphere approaching the force-free limit with particles supplied over the whole stellar surface. In Section \ref{sec:screening}, we show that a maximum in the dissipation occurs at an intermediate injection rate between the charge separated solutions and the force-free case (Figure \ref{fig:poynting}). Using dissipative models, \cite{Kala12} also found a maximum in dissipation at an intermediate conductivity. We can qualitatively associate the increasing conductivity of these models with the increasing injection rate in our simulations. \cite{Gruz13,Cont16b} have recently proposed \textit{weak pulsars}, that are magnetosphere configurations that present a larger dissipation than the force-free magnetosphere. These solutions are expected from a particle supply only at the neutron star surface, therefore they should be comparable to our simulations. We identified the solution with the highest dissipation at an intermediate $\mathcal{F}$. Considering also the previous results, we suggest that a \textit{weak pulsar} magnetosphere originates for these intermediate $\mathcal{F}$. A similar behavior is reported in \cite{Cerutti2015} for an aligned rotator and in \cite{Kalapotharakos2017b} for simulations that inject particles over the entire computational domain. 
\newline 
From the study of the macroscopic quantities and confirmed by the study of particle trajectories, we found that if particles are injected at the surface there are only a few regions of counter streaming particles. This is important because it was not clear if the currents in the magnetosphere were built of counter streaming species or not. For example, the photon-photon pair production in the current sheet \citep{Lyu96} would be inhibited if the electrons and positrons flow out in the same direction, as is the case in our simulation of plasma injection from the surface. Photon-photon pair production in the current sheet is implemented with simple prescriptions in other works (e.g \citealt{Chen2014}; \citealt{Philippov17}). We think that local simulations of this phenomenon in the pulsar current sheet and at the Y-point are needed to address this issue more carefully. However, there are also other mechanisms that could trigger the pair production, like considering other sources of photons, as was discussed for the outer gap \citep{ChiangRomani94}.
Another phenomenon impacted by this effect is the hypothesis of generation of the radio emission through the two stream instability (e.g. \citealt{usov2002} for a review). The two stream instability could still occur in the returning current region on the polar cap rim (note that for $45^{\circ}$ we found only one of the two branches to have counter streaming particles, see \ref{sec:macroscopic}) and below the pair formation front (e.g. \citealt{HarMus98}) that is not resolved by this simulation. 
\newline
When we look at the energetics of the most common particle trajectories (section \ref{sec:trajectories}) we see that the highest energy particles gain most of their energy close to the Y-point and they are outflowing positrons. Outflowing, energetic particles in the current sheet can produce light curves and spectra  \citep{Bra15, Kala14, Kala17a} that match well with those of the Fermi pulsars \citep{FermiPulsar}. In these works, the particle acceleration was allowed for outflowing particles only after the $R_{LC}$ where the Y-point is located and the particles were injected only at the surface. In Figure \ref{fig:screening}, we see that for magnetospheres that are far from the force-free limit, there can be some acceleration below the Y-point along the separatrix and above the polar cap. This probably indicates that the young $\gamma$-ray pulsars we selected for \cite{Bra15} have a magnetosphere close to the force-free limit. However, some millisecond $\gamma$-ray pulsars could have emission coming from these lower altitude gaps \citep{Tyrel14}, and their spectra would not suffer any magnetic pair attenuation because of the lower magnetic field. Other energetic particles are the electrons that are circling on the Y-point but they have lower energy than these outgoing positrons. These electrons, and the other particles that we showed crossing field lines, naturally gain pitch angles, thus breaking the ideal force-free limit. This makes them natural candidates for the non-thermal synchrotron emission observed at MeV energies and in the hard x-rays (e.g. \citealt{KuiHer15}) and it would explain the misalignment with the GeV emission that is observed in certain cases (e.g. \citealt{Marelli14}). Obviously the electrons that are circling around the Y-point are more promising candidates, but it is difficult to give final answers when nine orders of magnitude are squeezed into three. It would be extremely interesting to see at which energies this crossing of field lines happens for real pulsars with $\gamma_{max}\sim 10^9$, that would result in high energy particles with $\gamma\sim 10^7$ because of the radiation reaction. However, these kinds of works are helpful because they indicate a direction to follow and new hypotheses to be tested that did not emerge previously. Future missions looking at pulsars in the MeV band (e.g. \citealt{Astrogam}; \citealt{Compair}; \citealt{Adept}) could help unveil the mystery and constrain the models.
Looking at the particle trajectories showed also how a current structure close to the one of the force-free configuration can be sustained injecting particles only from the surface of the neutron star. Probably magnetospheres with pair production at the Y-point and/or in the current sheet or in other locations would settle on a different configuration, closer to the simulations where particles are injected everywhere. These scenarios will produce different signatures in the heating of the polar cap in addition to the heating generated by the pair production below the pair formation front (e.g. \citealt{HarMus01}). These signatures could be potentially observed and discriminated with NICER \citep{NICER,Ozel16}.   

\section{Conclusions}\label{sec:conclusions}
%%%%%%%%%%%%%%%%%%%%%%%%%%%%%%%%%%%%%%%%%%%%%%%%%%%%%%%%%%%%%%%%%%
We presented PIC simulations of the pulsar magnetosphere injecting particles only at the surface of the neutron star. We outlined the regime of our magnetosphere simulations, and we showed some properties of the solutions that are in between charge-separated solutions, and the force-free limit. Then we concentrated on a solution approaching the electromagnetic field structure of the force-free solution. We showed the different macroscopic quantities and compared them to a solution approaching the force-free limit where particles are injected everywhere in the domain. Following the differences, we studied the particle trajectories. The main findings are the behavior of the main flow of electrons and positrons in the magnetosphere and the complex mechanisms that sustain the current configuration.
We discussed both their theoretical and observational implications, underlining how comparing the same quantities in solutions obtained with different particle supply could be crucial for a theory-observation comparison. Future work could try to reproduce the injection in pulsar magnetospheres that are far from the force-free limit. In order to do this more realistically a more self-consistent injection mechanisms should be implemented, as for example injection dependent on the microphysics of the polar cap pair cascades. As we already mentioned in the text, more detailed studies of the pulsar current sheet \citep{DeVore2015} and Y-point would be helpful as well. 
%%%%%%%%%%%%%%%%%%%%%%%%%%%%%%%%%%%%%%%%%%%%%%%%%%%%%%%%%%%%%%%%%%%%%
%%%%%%%%%%%%%%%%%%%%%%%%%%%%%% ACKNOWLEDGEMENTS %%%%%%%%%%%%%%%%%%%%%
\acknowledgments 
%%%%%%%%%%%%%%%%%%%%%%%%%%%%%%%%%%%%%%%%%%%%%%%%%%%%%%%%%%%%%%%%%%%%%
This work is supported by the National Science Foundation under Grant No. AST-1616632, by the NASA Astrophysics Theory Program, by the NASA Astrophysics Data Analysis Program, and by Fermi Guest Investigator Program.
We want to acknowledge the support from the NASA High-End Computing (HEC) Program through the NASA Advanced Supercomputing (NAS) Facility at NASA Ames Research Center and NASA Center for Climate Simulation (NCCS) at NASA Goddard Space Flight Center.\newline
We thank the anonymous referee for useful suggestions that helped to improve the manuscript.\newline
Gabriele Brambilla wants to thank prof. Ioannis Contopoulos for the encouragment given at the \textit{The Physics of Pulsar Magnetospheres} workshop held at the NASA Goddard Space Flight Center in the June of 2016.  
%%%%%%%%%%%%%%%%%%%%%%%%%%%%%%%%%%%%%%%%%%%%%%%%%%%%%%%%%%%%%%%%%%%%%

%%%%%%%%%%%%%%%%%%%%%%%%%%%%%%%%%%%%%%%%%%%%%%%%%%%%%%%%%%%%%%%%%%%%%
% BIBLIOGRAPHY
%%%%%%%%%%%%%%%%%%%%%%%%%%%%%%%%%%%%%%%%%%%%%%%%%%%%%%%%%%%%%%%%%%%%%

%%%%%%%%%%%%%%%%%%%%%%%%%%%%%%%%%%%%%%%%%%%%%%%%%%%%%%%%%%%%%%%%%%%%%

\end{document}